\documentclass{jpp}

\usepackage{graphicx}
\usepackage{natbib}

\ifCUPmtlplainloaded \else
  \checkfont{eurm10}
  \iffontfound
    \IfFileExists{upmath.sty}
      {\typeout{^^JFound AMS Euler Roman fonts on the system,
                   using the 'upmath' package.^^J}%
       \usepackage{upmath}}
      {\typeout{^^JFound AMS Euler Roman fonts on the system, but you
                   dont seem to have the}%
       \typeout{'upmath' package installed. JPP.cls can take advantage
                 of these fonts, if you use 'upmath' package.^^J}%
      }
  \else
  \fi
\fi

\ifCUPmtlplainloaded \else
  \checkfont{msam10}
  \iffontfound
    \IfFileExists{amssymb.sty}
      {\typeout{^^JFound AMS Symbol fonts on the system, using the
                'amssymb' package.^^J}%
       \usepackage{amssymb}%

      }{}
  \fi
\fi

\ifCUPmtlplainloaded \else
  \IfFileExists{amsbsy.sty}
    {\typeout{^^JFound the 'amsbsy' package on the system, using it.^^J}%
     \usepackage{amsbsy}}
    {\providecommand\boldsymbol[1]{\mbox{\boldmath $##1$}}}
\fi

\newsavebox{\astrutbox}
\sbox{\astrutbox}{\rule[-5pt]{0pt}{20pt}}

\title[Optimum angle for side injection]{Optimum angle for side injection of electrons into linear plasma wakefields}

\author[K. V. Lotov]%
{K.\ns V.\ns L\ls O\ls T\ls O\ls V\thanks{Email address for correspondence: K.V.Lotov@inp.nsk.su}}

\affiliation{Budker Institute of Nuclear Physics SB RAS, Novosibirsk, 630090, Russia\\[\affilskip]
Novosibirsk State University, Novosibirsk, 630090, Russia}

\pubyear{2011}
\volume{65?}
\pagerange{???--???}
\date{?; revised ?; accepted ?. - To be entered by editorial office}
\begin{document}

\maketitle

\begin{abstract}
A unified model of electron penetration into linear plasma wakefields is formulated and studied. The optimum angle for side injection of electrons is found. At smaller angles, all electrons are reflected radially. At larger angles, electrons enter the wakefield with superfluous transverse momentum that is unfavorable for trapping. Separation of incident electrons into penetrated and reflected fractions occur in the outer region of the wakefield at some ``reflection'' radius that depends on the electron energy.
\end{abstract}


\section{Introduction}
Active studies of plasma wakefield acceleration initiated by \cite{Taj} for laser drivers and \cite{Chen} for electron ones have reached a point where quality of accelerated beams comes to the fore. Various injection techniques are now under investigation in search of reliable ways of producing low-emittance low-energy spread electron beams, as reviewed by \cite{RMP81-1229}. One promising technique is side injection of a low energy electron bunch to the wakefield at some small angle. This method was first proposed by \cite{PoP13-113102} to increase trapping efficiency by taking advantage of a specific wave nonlinearity. Later, injection at an angle was theoretically studied by \cite{PoP14-083101, LPB27-69} as a cure for ponderomotive scattering and deleterious effects on the vacuum-plasma transition. Recently, side injection has found application in proton driven wakefield acceleration \citep{elinj} as the means to save electrons from scattering at early stages of driver evolution in the plasma.

The theory of side injection is still a long way from completion. It is essentially two-dimensional. The one-dimensional Hamiltonian approach of \cite{PoP2-1432} helps to some extent to identify rough scalings, but quantitative results necessarily rely on simulations. In this paper we make a step toward creation of a two-dimensional trapping theory and formulate necessary conditions for side injected electrons to penetrate into the wakefield. We assume electrons are injected at a small angle $\alpha$ to the direction of driver propagation and show that there is an optimum value of this angle.

We first formulate the necessary mathematical framework for particle beam-driven wakes (section~\ref{pwfa}), as there is a universal field asymptotic in this case. In section~\ref{penetr} we analyze the equation obtained and find unified laws of electron penetration into the wakefield. In section~\ref{lwfa} we generalize the formalism to laser-driven wakes.

\section{Particle beam-driven wakes} \label{pwfa}
We use cylindrical coordinates $(r, \varphi, z)$ with the drive beam propagating along the $\bf z$ axis. The wakefield is assumed axisymmetric and unchanged in time in the driver reference frame. This assumption is justified if electron trapping occurs at a much shorter time scale than the time of driver evolution, which is typically the case. The phase velocity of the wakefield is $V_{w} \approx c$, the relativistic factor is $\Gamma_w = (1-V_w^2/c^2)^{-1/2}$, where $c$ is the speed of light.

\begin{figure}
  \centerline{\includegraphics{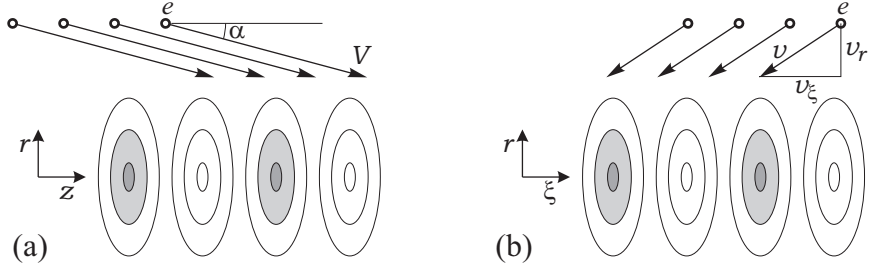}}
  \caption{The geometry of the problem in laboratory (\textit{a}) and co-moving (\textit{b}) coordinates.}
\label{fig1-geom}
\end{figure}
The geometry of the problem is shown in figure~\ref{fig1-geom}. Electrons initially form a stream of the velocity $V$ directed to axis. We use the co-moving coordinate $\xi=z-V_w t$ and do not Lorentz transform the fields into the moving frame. In new coordinates, electrons move backward and radially with velocity components
\begin{equation}\label{e1}
    v_\xi = V \cos \alpha - V_w, \qquad
    v_r = -V \sin \alpha \approx -\alpha c.
\end{equation}
As far as the side injection is usually aimed at further acceleration of electrons, we assume $V_w >  V$ and $v_\xi <0$.

Electron energies discussed in the context of side injection mostly fall in the range 3--15 MeV \citep{PoP14-083101, LPB27-69, elinj}, and usually the condition $\alpha \lesssim \Gamma^{-1}$ is fulfilled, where $\Gamma = (1-V^2/c^2)^{-1/2}$. As we show below, for the optimum angle the stronger condition $\alpha \Gamma \ll 1$ must be fulfilled, and we use this to neglect the angular dependence of the longitudinal velocity:
\begin{equation}\label{e2}
    v_\xi \approx c \left( 1 - \frac{1}{2\Gamma^2} \right) \left( 1 - \frac{\alpha^2}{2}  \right) - c \left( 1 - \frac{1}{2\Gamma_w^2} \right)
    \approx c \left( \frac{1}{2\Gamma_w^2} - \frac{1}{2\Gamma^2}\right) \approx V - V_w.
\end{equation}

The force exerted on an axially moving relativistic electron is the gradient of the wakefield potential energy $\Phi$. Side injected electrons either fall into the potential well of the wakefield, or are reflected radially by a potential hump. The initial energy $W_r$ of electron transverse motion is determined by the radial momentum $p_r \approx \alpha \Gamma m c$, where $m$ is the electron mass,
\begin{equation}\label{e3}
    W_r = \frac{p_r^2}{2 \Gamma m} \approx \frac{\alpha^2 \Gamma}{2} mc^2.
\end{equation}
For small $\alpha$, this energy contains the product of two small parameters ($\alpha$ and $\alpha \Gamma$) and is much smaller than height of any potential hump in wakefield structures of interest. Consequently, the choice of whether electron enters the wakefield or is reflected radially is made at large radii. Once an electron is trapped by the potential well, its radial velocity quickly increases, and the electron approaches the axis with little change in $\xi$-coordinate. Thus we are interested in potential behavior at large radii where important characteristics of trapping are determined.

For the particle driver of charge density $\rho(r,\xi) = \rho_b f(r) g(\xi)$ and the linearly responding plasma, the wakefield potential energy is \citep{PAcc20-171}
\begin{equation}\label{e4}
    \Phi (r, \xi) = - \frac{4 \pi \rho_b e}{k_p^2} R(r) Z(\xi),
\end{equation}
where
\begin{equation}\label{e5}
    R(r) = k_p^2 \int_0^r f(r') I_0(k_p r') K_0 (k_p r)\, r' \, dr' + k_p^2 \int_r^\infty f(r') I_0(k_p r) K_0 (k_p r')\, r' \, dr',
\end{equation}
\begin{equation}\label{e6}
    Z(\xi) = k_p \int_\xi^\infty g(\xi') \sin \left[ k_p (\xi'-\xi) \right] \, d\xi',
\end{equation}
$$
    k_p=\frac{\omega_p}{c}, \qquad \omega_p = \sqrt{\frac{4 \pi n_0 e^2}{m}},
$$
$n_0$ is the plasma density, $e>0$ is the elementary charge, $I_0$ and $K_0$ are zeroth order modified Bessel functions. At large radii and behind the driver, formulae (\ref{e4})--(\ref{e6}) take the universal form
\begin{equation}\label{e7}
    \Phi (r, \xi) = \Phi_0 \cos (k_p \xi + \phi_0) K_0(k_p r)
\end{equation}
with the amplitude $\Phi_0$ and phase $\phi_0$ determined by the individual driver shape. Setting $\phi_0=0$ by the choice of origin and using the asymptotic form of Bessel function $K_0$, we simplify the potential energy to
\begin{equation}\label{e8}
    \Phi (r, \xi) \approx \Phi_0 \cos (k_p \xi) \sqrt{\frac{\pi}{2}} \frac{\displaystyle e^{-k_p r}}{\sqrt{k_p r}}.
\end{equation}
Separation of incident electrons into trapped and reflected fractions occurs at some radius $r_0 \gg k_p^{-1}$ which we define later. Denote $\Phi_1 = \Phi (r_0, 0)$. Then, in the vicinity of $r_0$, we can write
\begin{equation}\label{e9}
    \Phi (r, \xi) \approx \Phi_1 \cos (k_p \xi) \, e^{-k_p (r-r_0)}.
\end{equation}
Equations of electron motion in the potential (\ref{e9}) are
\begin{equation}\label{e10}
    \frac{d p_r}{dt} = - \frac{\partial \Phi}{\partial r} = k_p \Phi_1 \cos (k_p \xi) \, e^{-k_p (r-r_0)},
\end{equation}
\begin{equation}\label{e11}
    \frac{d r}{dt} = \frac{p_r}{\Gamma m}, \qquad \frac{d \xi}{dt} = v_\xi.
\end{equation}
Here we neglect the change of total electron energy since it happens on much longer timescales.
Equations (\ref{e10})--(\ref{e11}) can be combined into one:
\begin{equation}\label{e12}
    \frac{d^2 r}{d t^2} = \frac{k_p \Phi_1}{\Gamma m} \cos [k_p (v_\xi t + \xi_0)] \, e^{-k_p (r-r_0)}.
\end{equation}
Introducing dimensionless variables
\begin{equation}\label{e13}
    \tilde t = -k_p v_\xi t, \qquad
    \tilde \xi_0 = k_p \xi_0, \qquad
    \tilde \xi = \tilde \xi_0 - \tilde t, \qquad
    \tilde x = k_p (r- r_0) - \ln \left(  \frac{\Phi_1}{\Gamma m v_\xi^2} \right),
\end{equation}
we rewrite equation (\ref{e12}) in a universal form:
\begin{equation}\label{e14}
    \frac{d^2 \tilde{x}}{d \tilde{t}^2} = \cos (\tilde{t} - \tilde{\xi}_0) \, e^{-\tilde{x}}.
\end{equation}
This equation must be solved with initial conditions corresponding to electron arrival from large $\tilde x$ with the negative dimensionless velocity
\begin{equation}\label{e15}
    \tilde v = \frac{d \tilde{x}}{d \tilde{t}} = -\frac{v_r}{v_\xi} = \frac{\alpha c}{V-V_w}.
\end{equation}

\section{Penetration of electrons into the wakefield}\label{penetr}

\begin{figure}
  \centerline{\includegraphics{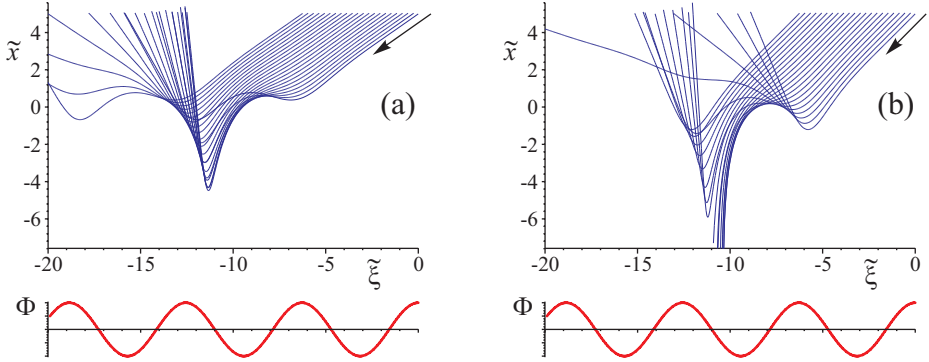}}
  \caption{Family of electron trajectories for $\tilde v = -0.7$ (\textit{a}) and $\tilde v = -1$ (\textit{b}). Lower graphs show the location of potential wells and humps.}
\label{fig2-traj}
\end{figure}

Equation (\ref{e14}) can be easily solved numerically for any values of initial velocity $\tilde v$ and phase $\tilde \xi_0$. For low initial velocities, all electrons are reflected by the outer region of the wakefield [figure~\ref{fig2-traj}(a)]. For larger $\tilde v$, there are two types of electron trajectories [figure~\ref{fig2-traj}(b)]. Depending on the phase, electrons either penetrate the wakefield, or are reflected.
The electrons entered into the wakefield are quickly accelerated radially and stick to some phase $\tilde \xi_t$. This is exactly the phase of the wakefield into which the injection of electrons occurs.

\begin{figure}
  \centerline{\includegraphics{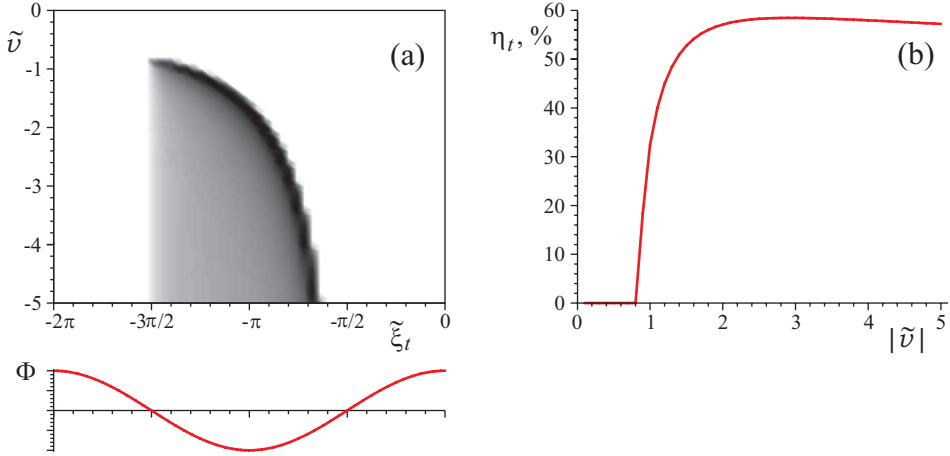}}
  \caption{Area of possible electron trapping on the plane $(\tilde \xi_t, \, \tilde v)$ (\textit{a}) and total penetration efficiency $\eta_t$ as the function of the initial velocity $\tilde v$ (\textit{b}). The lower graph shows the location of potential wells and humps.}
\label{fig3-map}
\end{figure}
\begin{figure}
  \centerline{\includegraphics{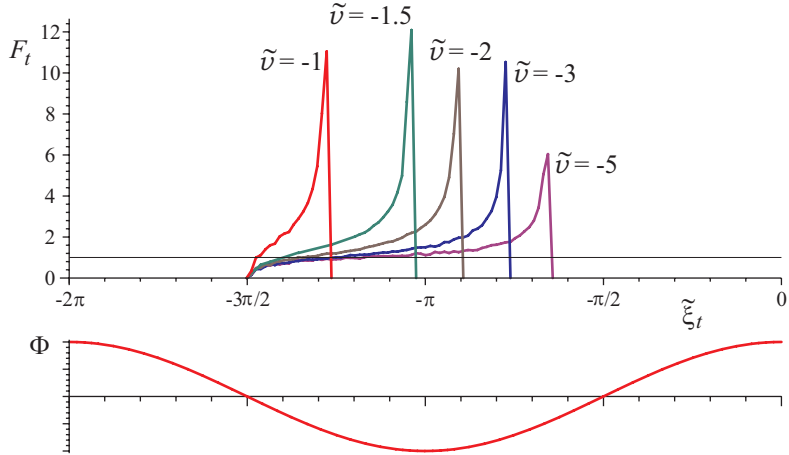}}
  \caption{Flux density $F_t$ of trapped electrons as the function of final phase $\tilde \xi_t$ for different values of the initial velocity $\tilde v$. Thin line is the flux density of incoming electrons. The lower graph shows the location of potential wells and humps.}
\label{fig4-cut}
\end{figure}

General picture of electron penetration is shown in figure~\ref{fig3-map}(a). Electrons cannot enter the wakefield at phases and initial velocities marked by the white color. The darker the color the greater the flux density $F_t$ of entered electrons, i.e. the number of electrons trapped in the unit interval of $\tilde \xi_t$ divided by the number of incoming electrons passing through a unit interval of $\tilde \xi$ at large radii. The total penetration efficiency $\eta_t$ (into all phases) is shown in figure~\ref{fig3-map}(b). The penetration threshold is observed at $\tilde v \approx -0.81$. For $|\tilde v| \gtrsim 1$, not only more than half the electrons penetrate into the wakefield, but also the  electron flux density increases several times (figure~\ref{fig4-cut}). In other words, injection of electrons preferably occurs into a certain phase of the wakefield, and this phase depends on the dimensionless radial electron velocity $\tilde v$.

Penetration of an electron into the wakefield does not mean this electron is trapped by the wakefield. The initial radial velocity of the electron can be too high, or the energy gained by the electron in near-axis regions can be insufficient for trapping. In these cases, the electron crosses the axis and leaves the wakefield on  the opposite side. The lower the initial radial velocity of the electron, the easier the electron to be trapped. Consequently, there is an optimum angle of electron injection corresponding to $|\tilde v| \sim 1$. It is just above the penetration threshold:
\begin{equation}\label{e16}
    \alpha_{\rm opt} \sim \frac{V_w - V}{c}.
\end{equation}
If the relativistic factor of the wave is much greater that that of the electron, $\Gamma_w \gg \Gamma$, then we obtain the engineering formula
\begin{equation}\label{e17}
    \alpha_{\rm opt} \sim \frac{1}{2 \Gamma_w^2}.
\end{equation}
For $\Gamma_w \sim \Gamma$, the optimum angle is even smaller. Estimate (\ref{e17}) justifies the condition $\alpha \Gamma \ll 1$ and formula (\ref{e2}).

As we see from figure~\ref{fig2-traj}, separation of incident electrons occurs at $\tilde x \approx 0$. Let us find the corresponding dimensional radius $r_0$. Substituting $\tilde x = 0$ and $r = r_0$ into definition of $\tilde x$ (\ref{e13}), we find
\begin{equation}\label{e18}
    \Phi (r_0, 0) = \Gamma m v_\xi^2.
\end{equation}
Using approximation (\ref{e8}) for $\Phi_1$, assuming $\Gamma_w \gg \Gamma$, and limiting ourselves to the logarithmic accuracy, we obtain a quick estimate for the reflection radius:
\begin{equation}\label{e19}
     r_0 \approx k_p^{-1} \ln \left(  \frac{\Phi_0 \Gamma^3}{m c^2}  \right).
\end{equation}

\section{Laser beam-driven wakes}\label{lwfa}
For laser drivers, there is no universal asymptotic form of the wakefield potential at large radii. The wakefield potential energy is determined by the vector potential of the driver $\bf A$ \citep{RMP81-1229}:
\begin{equation}\label{e20}
    \Phi (r, \xi) = - m c^2 k_p \int_\xi^\infty \frac{{\bf a}^2 (r, \xi')}{2}
    \sin \left[ k_p (\xi'-\xi) \right] \, d\xi', \qquad
    {\bf a} = \frac{e{\bf A}}{m c^2}.
\end{equation}
Usually $\Phi (r, \xi)$ follows the same radial dependence as ${\bf A}^2 (r, \xi)$ does. This dependence does not necessarily have an exponential asymptotic at large radii. Nevertheless, the results of section~\ref{penetr} are still applicable to laser driven fields though with a lower accuracy. We can approximate the potential energy in the vicinity of separation radius $r_0$ by the proper exponent function:
\begin{equation}\label{e21}
    \Phi (r, \xi) \approx \Phi_1 \cos (k_p \xi) \, e^{-\lambda (r-r_0)}.
\end{equation}
For example, for a Gaussian driver of the radius $\sigma_r$ we have
\begin{equation}\label{e4-3}
    {\bf a}^2 \propto e^{-r^2/\sigma_r^2}, \quad
    \Phi (r_0+\delta r, \xi) = \Phi_1 \cos (k_p \xi) \, e^{-2 r_0 \delta r / \sigma_r^2 - \delta r^2 / \sigma_r^2}, \quad
    \lambda = 2 r_0 / \sigma_r^2.
\end{equation}
As far as $\lambda \neq k_p$, equation (\ref{e12}), definition of $\tilde x$ (\ref{e13}), and the initial velocity (\ref{e15}) need to be modified:
\begin{equation}\label{e4-4}
    \frac{d^2 r}{d t^2} = \frac{\lambda \Phi_1}{\Gamma m} \cos [k_p (v_\xi t + \xi_0)] \, e^{-\lambda (r-r_0)}.
\end{equation}
\begin{equation}\label{e4-5}
    \tilde x = \lambda (r- r_0) - \ln \left(  \frac{\Phi_1 \lambda^2}{\Gamma m v_\xi^2 k_p^2} \right), \qquad
    \tilde v = \frac{d \tilde{x}}{d \tilde{t}} = -\frac{v_r \lambda}{v_\xi k_p}.
\end{equation}
In newly defined variables, equation (\ref{e14}) remains unchanged, as are the results of section~\ref{penetr} presented in figures~\ref{fig2-traj}--\ref{fig4-cut}. We need to modify only expression (\ref{e16}) for $\alpha_{\rm opt}$ and equation (\ref{e18}) for $r_0$:
\begin{equation}\label{e4-6}
    \alpha_{\rm opt} \sim \frac{(V_w - V) k_p}{c \lambda}, \qquad
    \Phi (r_0, 0) \lambda^2 (r_0)
    = \frac{1}{\Phi (r_0, 0)} \left( \frac{\partial \Phi (r_0, 0)}{\partial r_0} \right)^2
    = \Gamma m v_\xi^2 k_p^2.
\end{equation}

\section{Summary}
Let us summarize the main findings. There is an optimum angle for side injection of electrons into the plasma wakefield, as is given by formulae (\ref{e16}) or (\ref{e4-6}). At smaller angles, all electrons are reflected back. At larger angles, electrons enter the wakefield with superfluous transverse momentum that is unfavorable for trapping.

Penetration efficiency and the wakefield phase into which electrons enter are determined by the field behavior at large radii. In most cases, the wakefield there decays exponentially, and the process of electron entrance into the wakefield can be studied in a unified way.

Although formulated for linear wakefields, the results of the paper are also applicable to nonlinear wakes if the latter have a factorable field asymptotic of the form (\ref{e21}).

This work is supported by RFBR Grants 09-02-00594, 11-02-00563, and by the Russian Ministry of Education grants 2.1.1/3983 and 14.740.11.0053.

\end{document}